\documentclass[nofootinbib,apsspec,showpacs,twocolumn,groupedaddress,lengthcheck,preprintnumbers]{revtex4}
\usepackage{amsmath}
\usepackage{amssymb}
\usepackage{graphics}
\usepackage[dvips]{graphicx}
\usepackage{graphicx}
\usepackage[usenames]{color}

\def\simge{\mathrel{%
       \rlap{\raise 0.511ex \hbox{$>$}}{\lower 0.511ex \hbox{$\sim$}}}}
\def\simle{\mathrel{
       \rlap{\raise 0.511ex \hbox{$<$}}{\lower 0.511ex \hbox{$\sim$}}}}

\newcommand \beq{\begin{eqnarray}}
\newcommand \eeq{\end{eqnarray}}

\newcommand{\blueflag}[1]{{\color{black} #1}}

\usepackage[colorlinks=true,linktocpage=true,linkcolor=blue,citecolor=blue]{hyperref}

\begin{document}
\title{Evolution of Primordial Neutrino Helicities in Cosmic Gravitational Inhomogeneities}
\author{Gordon Baym and Jen-Chieh Peng}
\affiliation{\mbox{Illinois Center for Advanced Studies of the Universe}\\}
\affiliation{\mbox{
Department of Physics, University of Illinois, 1110
  W. Green Street, Urbana, IL 61801
  } \\
}

\date{\today}

\begin{abstract}
 
Relic neutrinos from the Big Bang decoupled from the hot plasma predominantly in helicity eigenstates.  Their
subsequent propagation through gravitational inhomogeneities of the Universe alters the helicities of both Dirac and Majorana neutrinos, thus providing an independent probe of the evolving universe.
 We determine here the probability that relic neutrinos flip their helicity, in terms of the spectrum of density inhomogeneities measured
in the Cosmic Microwave Background.  
 As we find, for Dirac neutrinos the gravitational helicity modifications are intermediate between the effects of 
magnetic fields if the neutrino magnetic moment is of the magnitude predicted in the Standard Model and the much larger effects if the
magnetic moment is of the scale consistent with the excess of low energy electron events seen by the XENON1T experiment.  We give succinct derivations, within general relativity, of the semi-classical response of a spinning particle to a weak gravitational field in an expanding universe,
and estimate the helicity modifications of neutrinos emitted by the Sun caused by the Sun's gravity.


\end{abstract}

\maketitle

\section{Introduction}

    The Cosmic Neutrino Background (C$\nu$B), analogous to the cosmic Microwave Background (CMB), carries invaluable independent information on the early universe \cite{dolgov,quigg,long,linholder}.   The primordial electron, muon, and tau neutrinos decoupled  in helicity eigenstates at 
temperatures $\sim$ MeV, much greater than neutrino masses, and cooled in the expanding universe to a present temperature $\sim 1.7 \times 10^{-4}$ eV.   Detection of the C$\nu$B,  a major experimental challenge, remains
an elusive goal.  The PTOLEMY experiment \cite{ptolemy} proposes to use inverse
tritium beta decay (ITBD) \cite{weinberg}, $\rm \nu_e + ^3H \to e^- +^3He$, to capture the relic neutrinos.  As the ITBD detection rate
depends on the helicity as well as the Dirac vs.\ Majorana nature of the
relic neutrinos \cite{long,numag}, a key question is to investigate how the
helicity of relic neutrinos evolve as they propagate through the Universe. 

  As first noted in Ref.~\cite{duda} a neutrino propagating in a gravitational field can develop an amplitude to have its helicity reversed; as
the neutrino trajectory is bent by a gravitational field, the bending of its spin lags the bending of the momentum \cite{silenko,dvornikov}.   A simple example is a finite mass neutrino with negative helicity shot straight upward from Earth at less than escape velocity; the neutrino will at a certain point reverse course and fall back down, but its spin direction will not be affected by the Earth's gravity (neglecting the Lense-Thirring effect from the Earth's rotation).  The result is that the neutrino returns with its momentum parallel to its spin, i.e., its helicity is flipped.  As another expample, the momentum of a non-relativistic neutrino in a circular orbit around a non-rotating gravitating point mass precesses by angle 2$\pi$ per orbit, while the spin precession is a relativistic correction \cite{schiff}.  Thus non-relativistically the neutrino helicity oscillates between negative and positive helicity in half an orbit.   
    
  A second effect that can modify the helicity of Dirac, but not Majorana, 
neutrinos arises from their expected magnetic 
moment \cite{marciano,benlee,fujikawa,lynn,s-w,bell-Dirac,bell,dolgov,gs}, 
which is diagonal in the mass eigenstate basis. Majorana neutrinos 
can only have non-diagonal transition magnetic moments between different mass 
eigenstates.  As a Dirac neutrino propagates through astrophysical 
magnetic fields, from cosmic to galactic to magnetic fields in supernovae 
and neutron stars, its spin precesses and its helicity is modified.  
As we discussed, the helicity modification is sensitive both to the
neutrino magnetic moment and to the characteristics of the magnetic
fields ~\cite{numag}.  In estimating the helicity flipping probability
for relic neutrinos in both cosmic and galactic magnetic fields, we found 
that even a neutrino magnetic moment well below the value suggested by
the XENON1T experiment could significantly affect the helicities of relic neutrinos,
and their detection rate via the ITBD reaction~\cite{numag}.    
  
    We focus here on the gravitational effect on the helicities of 
relic neutrinos as they propagate from the time of decoupling in the early 
universe, of order one second after the Big Bang, to the present. 
Owing to the charged current interaction for 
$\nu_e$ and $\bar \nu_e$, the reaction
cross sections for electron neutrinos are larger
than for muon and tau neutrinos. An immediate consequence is that electron neutrinos decouple from the plasma of the early universe at a later time and at a lower temperature than muon and tau neutrinos.
As estimated in Ref.~\cite{dolgov}, $\nu_\tau$ and $\nu_\mu$ freeze
out at temperature $T_\mu\sim  1.5$ MeV, while $\nu_e$ freeze out
at temperature $T_e\sim 1.3$ MeV. However, the temperature differences
at freezeout do not effect the present temperature, $T_{\nu 0}= 1.945 \pm 0.001$K = $(1.676\pm 0.001)\times 10^{-4}$ eV, of the various neutrino species (a
factor $(11/4)^{1/3}$ smaller than that of the cosmic
microwave background).

   Relic neutrinos are produced in flavor eigenstates, a 
coherent sum of neutrino mass eigenstates,  and in wave packets whose structure is determined effectively by the electrons and positrons scattering with the $\nu$ and  $\bar\nu$.  The wave packets are limited in size by electron mean free paths at the time of decoupling; as calculated in Ref.~\cite{henning},
a characteristic electron mean free path is of order $1/\alpha^2 T$ to within logarithmic corrections, where $\alpha = e^2/4\pi$; thus at $T\sim$ 1 MeV, the electron mean free path is of order $10^6-10^7$ fm. 

  The wave packets
of flavor eigenstates quickly disperse into three effectively decoherent  wavepackets each with a given mass, owing to their velocity differences. The velocity dispersion of the mass eigenstates of a relativistic neutrino with momentum $p$ at decoupling is $\delta v/c \simeq \frac12 \Delta m^2 /p^2$, where $\Delta m^2$ is the characteristic neutrino mass-squared splitting \cite{masses}.   With $\Delta m^2$ on the characteristic scale of $10^{-4}$ eV$^2$, the velocity dispersion for $p \sim$ 1 MeV is
$\sim 1.5 \times 10^{-6}$ cm/sec;  thus in the first second alone after neutrinos are decoupled, dispersion would spread the mass components some $10^7$ fm, at least on the scale of the wave packets in which the neutrinos are produced.  The decrease of $p$ in time only increases the velocity dispersion.    By contrast, the velocity dispersion within a wave packet of definite mass, $ \sim   (\delta p/p) m^2 /p^2$, is much smaller, since $\delta p$ within a wavepacket is small compared with the packet's mean momentum $p$. 

   At freezeout the neutrinos are left in a relativistic thermal distribution,
\beq
   f(p) = \frac{1}{e^{p/T}+1},
   \label{distribution}
\eeq
where $p$ is the neutrino momentum and $T$ the temperature;  this distribution is 
maintained throughout the evolution of the universe, even though neutrinos in at least two of the three mass states are non-relativistic at present. 

  In the following Section,~\ref{spinrotation}, we lay out the basic physics of momentum spin rotation by a weak gravitational potential, giving self-contained semiclassical derivations from general relativity of the effects in Appendix A.    Then in Sec.~\ref{expansion} we calculate the net momentum rotation of primordial neutrinos propagating through the gravitational inhomogeneities of the expanding universe -- the gravitational lensing of the C$\nu$B -- and the net helicity changes the neutrinos undergo.  As a related application we estimate in Sec.~\ref{solarneutrinos} the expected helicity rotation of solar neutrinos caused by their gravitational interaction with the Sun itself. In the concluding Section, \ref{conclusion}, we compare the gravitational bending with the rotation of neutrino spins owing to a finite neutrino magnetic moment, estimated earlier \cite{numag}.   Appendix B provides a detailed derivation of the bending of neutrinos emitted from compact spherical objects such as the Sun, neutron stars, and supernovae.  We work in units with $\hbar = c =1$. 
  
\section{Spin rotation in a weak gravitational potential
\label{spinrotation}}

 When a particle of mass $m$ and velocity $\vec v$ propagates through a weak gravitational potential $\Phi$ its direction of momentum, $\hat p$, bends at a rate
\beq
     \frac{d\hat p}{dt}\Big|_\perp = -\left(v + \frac{1}{v}\right)\vec \nabla_\perp \Phi,
      \label{mombend}
\eeq
where the gradient is taken perpendicular to the direction of momentum.   We measure the spin precession in $\Phi$ in terms of the particle spin 
$\vec S$ in the particle's local Lorentz rest frame, reached by a Lorentz boost without rotation.  The spin precesses at the slower 
rate~\cite{voronov,silenko}, 
\beq
   \frac{d\vec S}{dt}\Big|_\perp   = -\frac{2\gamma+1}{\gamma+1}\vec S\cdot \vec v\,\, \vec\nabla_\perp\Phi,
    \label{spinbend}
\eeq
where  $\gamma = 1/\sqrt{1-v^2}$ is the usual Lorentz factor.
These results are derived in Appendix \ref{gr}, including the expansion of the universe.    In a helicity eigenstate $\hat S\cdot\hat p  = \hat S\cdot\hat v = h = \pm 1$,
one has equivalently,
\beq
    \left[h\frac{d\hat S}{dt} - \frac{d\hat p}{dt}\right]_\perp  = \frac{m}{p} \vec\nabla_\perp\Phi.
    \label{deltaspinbend}
\eeq

  As a consequence of the spin lagging the momentum, the helicity of the particle is rotated by gravitational fields.   For total angular bend 
$\delta\theta_p$ of the momentum, determined by Eq.~(\ref{mombend}), the angular bend, $\delta \theta$, of the spin with respect to the momentum is thus 
\beq
 \delta\theta = \delta\theta_{s}-\delta\theta_p = - \frac{\delta\theta_p}{\gamma(1+v^2)}.
    \label{lag}
\eeq
where $\delta\theta_s$ is the bending angle of the spin, calculated from Eq.~(\ref{spinbend}).

\subsection{Helicity change in passing a distant point mass}

   A simple application is the deflection of a relativistic spinning particle passing a distant point mass $M$.    Integrating the transverse acceleration (\ref{mombend}) over the particle trajectory from $t= -\infty$ to $\infty$ one finds the expected deflection,
\beq
  \Delta\theta_p = \frac{2MG}{bv^2}(1+v^2),
  \label{Deltatheta}
\eeq
where $G$ is the Newtonian gravitational constant, and $b$ is the impact parameter.  
(For $v=1$ this is the Einstein weak field light-bending result.)   The spin axis precesses by the smaller amount, 
\beq
   \Delta \theta_{s}  = \frac{2MG}{b}\frac{2\gamma+1}{\gamma+1},
\eeq
and the angular change of the spin axis with respect to the momentum axis is
\beq
   \Delta \theta= - \frac{2MG}{b \gamma v^2}.
   \label{thetaM}
\eeq

 In the fully relativistic limit, the spin tracks the momentum, leading to no change in the particle helicity.  On the other hand, in the non-relativistic limit the spin rotates negligibly compared with the bending of its momentum, and thus a change in direction of the momentum leads to a change in particle helicity.    
For spin rotation with respect to the momentum  by angle $\theta$ from an initial helicity state, the helicity changes from $\pm 1 $ to  $\pm \cos\theta $, and the probability of observing the spin flipped to the opposite direction, which is half the magnitude of the change in helicity, is then $P_f = \sin^2(\theta/2)$.

\section{Integrating over the expansion of the universe
   \label{expansion}}
    
       We now calculate the momentum bendings, and then spin rotations, as neutrinos propagate past the density fluctuations in the early universe.  To take into account the expansion of the universe, we work in terms of the   
standard Friedman-Robertson-Walker metric, 
\beq
    ds^2 = a(u)^2[-du^2 + d\vec x\,^2].
    \label{ga}
\eeq
Here $u$ is the conformal time, related to coordinate time, $t$, by
$dt = a(u)\,du$, with the metric in homogeneous space; and $\vec x$ are the comoving spatial coordinates, related to the usual spatial coordinates, $\vec r$, by $d\vec r = a(u) d\vec x$.   We take $a(u)=1$ at present.
   
      In the presence of small energy density fluctuations, $\rho(x) = \bar \rho + \delta\rho(x)$, with $\bar \rho$ the spatially uniform average density, the metric (\ref{ga}) becomes \cite{hartle}
\beq
     ds^2 = a(u)^2[-(1+2\Phi) du^2 + (1-2\Phi)d\vec x\,^2],
     \label{metricphi}
\eeq
where the scalar potential $\Phi$ is given in terms of the density fluctuations by
\beq
    \nabla_x^2\Phi = 4\pi G\left(\delta\rho(\vec x\,)+3\delta P(\vec x\,)\right) a(u)^2,
    \label{phirho}
\eeq
with $\delta P$ is the variation of the pressure from uniformity, and   $a^{-1}\nabla_x$ the gradient with respect to $\vec r$.  

  In the matter-dominated era (denoted by $\cal M$), the pressure term can be neglected, and (\ref{phirho}) becomes the familiar Newtonian equation.   Furthermore in this era linear perturbation theory \cite{dodelson} implies that 
\beq
   \delta(\vec x\,) \equiv \delta\rho(\vec x\,)/\bar\rho
\eeq   
grows as $a$, where $\bar\rho$ is the average density; thus since $\bar\rho$ scales as $1/a^3$, we see immediately that  $\delta \rho(\vec x\,)$ scales as $a^{-2}$ and thus $\nabla_x^2\Phi(\vec x)$ and $\Phi(\vec x)$ as functions of $\vec x$ are constant in time.   

    In the radiation-dominated era (denoted by $\cal R$),  $\Phi(\vec x)$  as a function of $x$ is also constant in time, since in this era linear perturbation theory implies that $\delta$ grows rather as $a^2$ at large scales, while $\bar\rho$ and $\bar P$ scale as $1/a^4$.  Furthermore the pressure fluctuations in this era are simply 1/3 of the density fluctuations, so that $\nabla_x^2\Phi =  8\pi G a^2  \bar\rho(x) \delta(\vec x\,) $. 

   To calculate the angular changes in the trajectory of a neutrino, we neglect the neutrino mass at this point for simplicity.  Then Eq.~(\ref{mombend}) gives a total angular change $ - 2\int d\ell \, \nabla_{x\perp} \Phi(\vec x)$,
where $\ell$ is the comoving length along the path.   To lowest order the integral is along the straight path of the neutrino, parametrized in the absence of density fluctuations
by the coordinate $x_3$.   The average of the square of the angular deflection of the particle trajectory is then
 \beq
     \langle (\Delta\theta_p)^2\rangle = 4\int dx_3 dx'_3 \vec\nabla_{x\perp}\cdot\vec\nabla_{x'\perp}\langle \Phi(x_3) \Phi(x_3')\rangle,
     \label{13}
 \eeq
where
 \beq
  \langle \Phi(\vec x) \Phi(\vec x')\rangle =  \int \frac{d^3k}{(2\pi)^3} e^{i\vec k\cdot(\vec x-\vec x\,')} \Psi(k)
\eeq
is the spatially isotropic, (conformal) time-independent auto-correlation function of the gravitational perturbations; the vectors $\vec k$ are comoving. 

Then
\beq
     \langle (\Delta\theta_p)^2\rangle = 4\int dx_3 dx'_3 \int \frac{d^3k}{(2\pi)^3} e^{i k_3(x_3-x_3')} k_\perp^2 \Psi(k). \nonumber\\
     \label{13}
\eeq 
The integration over $x_3'$ essentially gives $2\pi\delta(k_3)$, so that 
\beq
     \langle (\Delta\theta_p)^2\rangle =  \frac{2}{\pi} \int du \int dk_\perp k_\perp^3\Psi(k_\perp),
          \label{13}
\eeq 
where  $x_3 = u$ along the trajectory of the neutrino.  

  The spectral function $\Psi(k)$ is directly related to the spectral function of the density correlation function,
\beq
   \langle \delta(\vec x\,)\delta(\vec x\,')\rangle = \int \frac{d^3k}{(2\pi)^3} e^{i\vec k\cdot(\vec x-\vec x\,')} P(k),
   \label{rhoB}
\eeq
by 
\beq
    \Psi(k) =   (4\pi G \bar \rho a^2)^2  \zeta \frac{P(k)}{k^4},
\eeq
with $\zeta $ = 1 in $\cal M$, and 4 in $\cal R$ where $\delta P= \delta\rho/3$.

  The spectral function $P(k)$  (with dimensions of volume) depends on the magnitude of $\vec k$ and the time.   Its general structure \cite{Planck} is an approximately Harrison-Zel'dovich long wavelength linear growth in $k$ below a maximum at wavevector $k_H$; for $k>k_H$, $P(k)$ falls roughly as $k^{-\nu}$ with $\nu>0$.    For $k$ below $k_H$, $P(k)$ scales in $\cal M$ as $a^2$ (even beyond the peak at $k_H$),  
and as $a^4$ in $\cal R$.   In terms of $P(k)$ (with the subscript $\perp$ on the integration variable dropped),   
\beq
     \langle (\Delta\theta_p)^2\rangle = 32\pi \zeta \int du (G\bar \rho a^2)^2 \int\frac{dk}{k} P(k).
     \label{13}
 \eeq

 The angular bending of the neutrino trajectories and modification of the helicity are largest in the matter-dominated era, on which we now focus. \blueflag{ We include  dark energy, which affects the cosmological expansion after redshifts of order 1/2.
 The relation between the scale factor and the conformal time is determined by
\beq
    \frac{da}{du} = \sqrt{\frac{8\pi G \bar \rho(a) a^4}{3}} = H_0 \sqrt{\Omega_M a + \Omega_V a^4},
    \label{dadude}
\eeq
where $\bar \rho(a) = \rho_M/a^3 + \rho_V$, with $\rho_M/\rho_c \equiv \Omega_M \simeq 0.32 $ the present average mass fraction (including dark matter) in the universe, $\rho_V/ \rho_c \equiv \Omega_V \simeq 0.68$ the dark energy fraction, and $\rho_c$ the present critical closure density;  
$H_0 = \sqrt{8\pi G \rho_c/3}$ is the present Hubble constant \cite{frieman,Planck6}.

    With $P_0(k) = P(k)/a^2$, the angular deviations produced in propagation from matter-radiation equality (where $a(t_{eq}) \equiv a_{eq}\sim 0.8\times 10^{-4}$) to now are given by
  \beq
    \langle (\Delta\theta_p)^2\rangle  \simeq  \frac{9}{2\pi}H_0^4 {\cal P} \int_{u_{eq}}^{u_0} du (\Omega_M+\Omega_Va^3)^2,   
     \label{dtheta11a}
\eeq
where ${\cal P} \equiv \int_0^\infty (dk/k) P_0(k)$.   Numerical integration of the Planck collaboration data \cite{Planck} -- Fig.~19, yields
${\cal P} \simeq 7.25 \times 10^4$ (Mpc/h)$^3$.

   Using $a$ as the independent integration variable in evaluating the rotation angles, we find 
\beq
     \langle (\Delta\theta_p)^2\rangle &=& \frac{9}{2\pi} {\cal P}H_0^3  \int_{a_{eq}}^1 \frac{da}{a^2}\left(\Omega_M a +\Omega_V a^4\right)^{3/2}.
          \label{detheta12a}
\eeq   
The $a$ integral is approximately 0.56.  In addition ${\cal P}H_0^3 \simeq 2.69 \times 10^{-6}$ (independent of the Hubble parameter $h$), and thus 
\beq
    \langle (\Delta\theta_p)^2\rangle \simeq 2.2 \times 10^{-6}.   
    \label{thetamdea}
\eeq
}

    This result indicates that gravitational lensing of the CMB would be \blueflag{$\sim$ 5.1 arcmin,} within a factor of two of the value $\sim$ 2.7 arcmin from more precise calculations,\footnote{Owing to reionization of intergalactic H atoms below redshift $z\sim 10$ and subsequent photon-electron scattering,  the lensing of the CMB is most efficient at lower redshift.    (Neutrino lensing does not experience such restrictions; the weak electron-neutrino scattering after reionization is insignificant in comparison.)   For example, integration over a sharply limited range of $z< 6$ in Eq.~(\ref{detheta12a}) reduces the mean bending angle to $\sim$ 3.9 arcmin.}  e.g., \cite{tonytony}.

\begin{figure}
\includegraphics*[width=8.5cm]{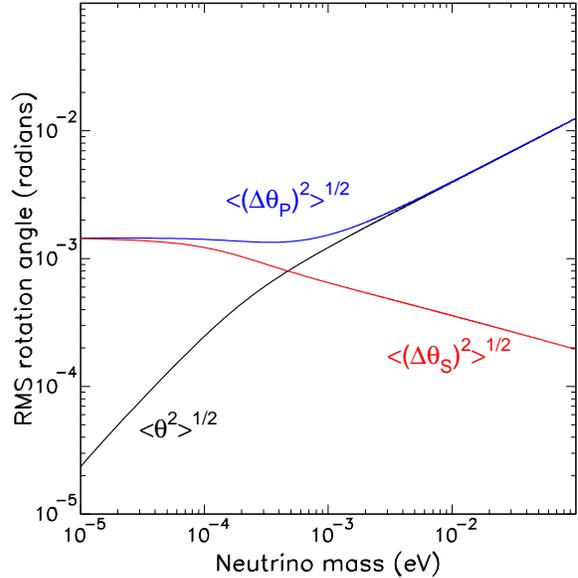}
\caption{The root mean square bending angles of the neutrino momentum $\sqrt{\langle (\Delta\theta_p)^2\rangle}$, spin $\sqrt{\langle (\Delta\theta_s)^2\rangle}$,  and the bending of the spin with respect to the momentum $\sqrt{\langle \theta^2\rangle}$, Eq.~(\ref{spinbend10}), in the matter-dominated era, as functions of the neutrino mass.   All curves are calculated for the neutrino momenta equal to the present neutrino temperature. 
The contribution to the bending angles from the radiation-dominated era  is negligible. }
 \label{nugrav_f}
\end{figure}

    We now consider the effect of the neutrino mass, which is significant only in $\cal M$.    For finite mass, the integration over $u$ in Eq.~(\ref{13}) now becomes
   \beq
   \frac14\int_{u_{eq}}^{u_0}du \,v(u)\left(v(u)+\frac1{v(u)}\right)^2,
   \label{18}
\eeq
as one sees from Eq.~(\ref{mombend}), with $dx_3 = v(u) du$.   \blueflag{This modification leads to
 \beq
   \langle (\Delta\theta_p)^2\rangle  &=&\frac{9}{8\pi} {\cal P}H_0^3  \int_{a_{eq}}^1 \frac{da}{a^2}\left(\Omega_M a +\Omega_V a^4\right)^{3/2} \nonumber\\
     && \hspace{48pt} \times v(a)\left(v(a)+\frac1{v(a)}\right)^2.
   \label{mombend10a}
 \eeq
The velocity of a neutrino of momentum $p_0$ at present, and thus with a comoving momentum $p= p_0/a$,  is $v(a) = 1/\sqrt{1+m_\nu^2 a^2/p_0^2}$. }  The root mean square bending angle, $\sqrt{\langle(\Delta \theta_p)^2\rangle}$, is shown in Fig.~\ref{nugrav_f} as a function of the neutrino mass.

    In the limit of a very slow neutrino, $p_0/m_\nu \ll 1$, the integral in Eq.~(\ref{mombend10a}) is $\simeq 0.3 m_\nu/p_0$, and we find
\blueflag{
\beq
    \langle (\Delta\theta_p)^2\rangle \simeq \frac{2.7}{8\pi} {\cal P}H_ 0^3  \frac{m_\nu}{p_0};
    \label{thetamde}
\eeq
}
the bending of a non-relativistic neutrino is larger, as one sees in Fig.~\ref{nugrav_f}, than the bending of a relativistic neutrino.

   In the radiation-dominated era, from the time of neutrino decoupling, $t_d \sim$ 1 s, to matter-radiation equality, the scale factor is linear in conformal time, $a(u) = (8\pi G\bar\rho a^4/3)^{1/2} u$, and thus from Eq.~(\ref{13}), 
\beq
  \langle (\Delta\theta_p)^2\rangle = \frac{18}{\pi} \frac{a_{eq}^4}{u_{eq}^4}  \int^{u_{eq}}_{u_d}  \frac{du}{a(u)^4}  \int\frac{dk}{k}P(k,u).
   \label{thetardeu}
\eeq
Density fluctuations grow in $\cal R$ as $a^2$, and thus $P(k)$ grows as $a^4$ outside the horizon scale.  The horizon grows as $t \sim a^2$ so that the physical wavevector of the horizon decreases as $1/a^2$ and the comoving wavevector decreases as $1/a$. This implies that the maximum, $P(k_H)$, of $P(k)$ for comoving $k$ grows as $a^3$, until matter-radiation equilibrium, after which it grows as $a^2$.   Since $\int dk P(k)/k$ is essentially proportional to $P(k_H)$, we infer,
\beq
    \int \frac{dk}{k}P(k,u) &\simeq& \frac{a(u)^3}{a_{eq}^3}  \int \frac{dk}{k}P(k,u_{eq})\nonumber\\
       &&\simeq \frac{a(u)^3}{a_{eq}}  \int \frac{dk}{k}P_0(k).
\eeq
With (\ref{thetardeu}),
\beq
     \langle (\Delta\theta_p)^2\rangle 
       &\simeq& \frac{18}{\pi} \frac{a_{eq}^{2}}{u_{eq}^3} \ln\left(\frac{a_{eq}}{a_d}\right)  \int \frac{dk}{k}P_0(k), \nonumber\\
       &\sim&\blueflag{a_{eq}^{1/2}  \ln\left(\frac{a_{eq}}{a_d}\right) {\cal P}H_0^3},  
              \eeq
where $a(u_d) \equiv a_d \sim 2.3\times 10^{-10}$, and we scale to the present, \blueflag{writing $u_{eq} \sim a_{eq}^{1/2}/H_0$.}  
The squared angular bending of momentum in the radiation-dominated era is thus of order \blueflag{a few} percent of  that in the matter-dominated era, \blueflag{ Eq.~(\ref{detheta12a})}.   
 
\blueflag{
  The spin axis rotates away from the momentum axis only in the matter dominated regime, where the finite neutrino mass can play a role.   To estimate the rotation of the spin itself, we replace according to  Eq.~(\ref{spinbend}), the factor $(v+1/v)$ by $v(2\gamma+1)/(\gamma+1)$ in Eq.~(\ref{mombend10a}), so that 
  \beq
     \langle (\Delta\theta_s)^2\rangle  &=& \frac{9}{8\pi} {\cal P} H_0^3 \int_0^1 \frac{da}{a^2} \left(\Omega_M a +\Omega_V a^4\right)^{3/2}  \nonumber\\    && \hspace{60pt} \times v^3\left(\frac{2\gamma+1}{\gamma+1}\right)^2 .
   \label{mombend12}
 \eeq
Similarly the probability of spin rotation away from a pure helicity state, is, according to Eqs.~(\ref{mombend}) and (\ref{lag}), given by Eq.~(\ref{mombend10a}) with the factor $(v+1/v)$ by $1/\gamma v =m_\nu/p$,  
\beq
     \langle \theta^2\rangle &=& \frac{9}{8\pi} {\cal P}H_0^3  \int_0^1 \frac{da}{a^2}\left(\Omega_M a +\Omega_V a^4\right)^{3/2}   \left(\frac{1}{v}-v\right), 
     \nonumber\\
          \label{spinbend10}
\eeq     
where   
\beq   
   \left(\frac{1}{v}-v\right) =  \frac{m^2a^2}{p_0\sqrt{p_0^2 + m^2 a^2}}. 
\eeq

 \begin{figure}
\includegraphics*[width=8.5cm]{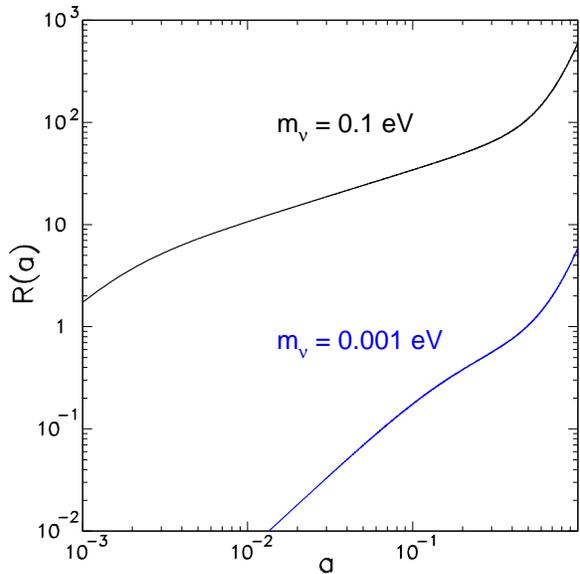}
\caption{\blueflag{ The integrand $R(a)$ in of the $a$ integral in Eq.~(\ref{spinbend10}), showing the dependence of the 
root mean square bending angle of the neutrino spin relative to the
momentum as a function
of the scale factor $a$,  for two neutrino masses, and momentum equal to the present neutrino temperature. }} 
 \label{nugrav_fig2}
\end{figure}
 
     Figure~\ref{nugrav_f} shows the bending of the momentum, Eq.~(\ref{mombend10a}), the bending of the spin, calculated using Eq.~(\ref{mombend12}), and the bending of the spin axis with respect to the momentum axis, Eq.~(\ref{spinbend10}), as a function the mass of the neutrino, for the neutrino momentum equal to the temperature.  
Similarly  Fig.~\ref{nugrav_fig2} shows the root mean square bending angle of the spin with respect to the momentum as a function of the scale factor $a$, for two representative neutrino masses.  As this figure shows, the onset of the role of dark energy in the expansion of the universe leads to a relative increase in the bending in recent epochs, $a\gtrsim 0.3$.  

}

  The equality of the spin rotation with respect to the momentum and the momentum rotation for a non-relativistic neutrino, seen in Fig.~\ref{nugrav_f}, is simply a consequence of the absence of spin rotation of a non-relativistic neutrino in a gravitational field;  for a relativistic neutrino, 
$\langle \theta^2 \rangle$ is suppressed by a factor $(m_\nu^2/2 p_0)^2$ compared with the momentum bending (\ref{thetamdea}).    To put the scale of bending in context, we note from Eq.~(\ref{thetaM}) that the spin rotation of a marginally non-relativistic neutrino ($p\sim m_\nu$) is of order that a neutrino would experience in passing a solar mass neutron star at a distance $\lesssim 10^4$ km.

\section{Helicity changes of solar neutrinos
\label{solarneutrinos}}

    A related application of helicity rotation by gravitational fields is the spin rotation of solar neutrinos in the gravitational fields of the Sun.  
To estimate the effects, we consider neutrinos emitted in the $z$-direction, focussing first on those emitted at a given transverse distance, $b$, from the $z$-axis, and distance $r_0$ from the center of the star.  Since emission at $-b$ leads to the same helicity change as $b$, and there is no coherence between emission from the points $\pm |b|$, we may take $b>0$ throughout.    Then the relative bending of the spin and momentum of these neutrinos is, from Eq.~(\ref{deltaspinbend}), given by
\beq
   \gamma v^2 \theta(b,r_0)  &=& \int_{z_0}^\infty dz \, \nabla_y \Phi(r) = -b\int_{z_0}^\infty dz \, \frac{G M(r)}{r^3}, \nonumber\\      
    \label{solbend}
\eeq 
where $M(r)$ is the stellar mass interior to radius $r$, and $z_0 = \pm \sqrt{r_0^2-b^2}$, with $z$ measured from the center of the star.   The dependence on the neutrino mass is entirely through the velocity dependent factor, $1/\gamma v^2$.

   Owing to the spherical symmetry of the Sun, the average bending of the neutrinos beginning at the two values of $z_0$ is just the same as if the neutrinos started from $z_0=0$.  Thus, in calculating the average helicity bending angle,  we can replace the lower limit in the integral by 0; the average is independent of $r_0$.   Averaging as well over the solar volume, weighted by $p_\nu (r)$, the normalized distribution of neutrino production in the Sun, we derive, as detailed in Appendix B, the average bending angle 
\beq
  \langle\theta\rangle  &=& - \frac{G}{\gamma v^2}\int_0^{R_\odot} 4\pi r_0 dr_0 p_\nu(r_0) \int_0^\infty dr\frac{M(r)}{r^2} f(r,r_0), \nonumber\\
 \label{solbend3}
\eeq
where
\beq
  f(r,r_0) = \Theta(r_0-r) r W(r_0/r)  + \Theta(r-r_0) r_0 W(r/r_0)  \nonumber\\
\eeq 
with the elliptic integral
\beq
   W(\xi) = \int_0^1 dx\frac{\sqrt{1-x^2}}{\sqrt{\xi^2-1+x^2}}, \quad \xi>1. 
\eeq

    Equation~(\ref{solbend3}) is a convenient starting point for integrating numerically over the empirical mass distribution $M(r)$ and neutrino emissivity distribution $p_\nu(r)$ of the Sun; using solar model distributions \cite{Bahcall} we find  
\begin{equation}
\langle \theta \rangle = -\frac{1.54}{\gamma v^2} \frac{GM}{R}  
\label{numerical}
\end{equation}
For a uniform mass density $\rho(r)$ and uniform $p_\nu(r)$, the prefactor becomes 0.76.

\begin{figure}[hbp]
\begin{center}
\includegraphics[width=8.0cm]{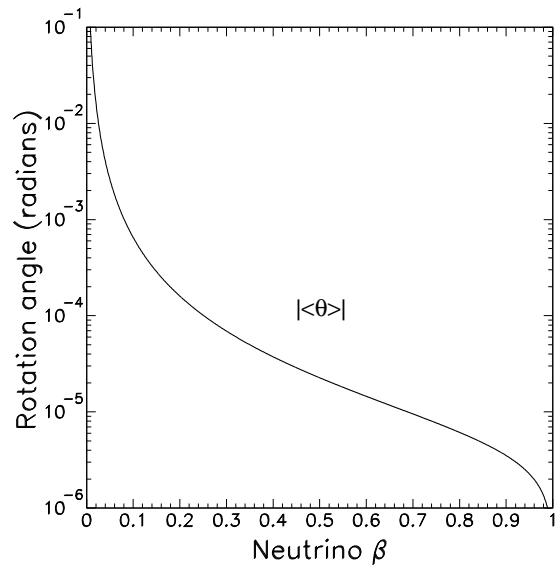}
\end{center}
\caption[*]{\baselineskip 1pt
The mean helicity rotation angle $|\langle \theta \rangle|$
for solar neutrinos as a function
of the neutrino $\beta=v/c$.} 
\label{solarnu}
\end{figure}

   As seen in Fig.~\ref{solarnu} the helicity bending angle $|\langle \theta\rangle|$ of non-relativistic solar neutrinos is sizable; however, only a tiny fraction of solar
neutrinos are non-relativistic.  On the other hand, heavy particles with non-zero spin, such as dark photons, emitted
from the Sun would have their helicities significantly modified
by the Sun's gravitational field.  How such a helicity rotation of dark photon could be observed remains an interesting question.

    To understand the magnitude of the helicity angle bending from the Sun,  we note that the average emission radius of neutrinos, 
$\langle r_0 \rangle =  \int d^3r\, r\, p_\nu(r)$ is $\simeq 0.11 R_\odot$, and thus $b\ll R_\odot$.   Since $b=r_0\sin\omega$, where
$\omega$ is the polar angle, the average value of $b$ is $\pi\langle r_0 \rangle/4$.  We can thus replace the $z$ integral in Eq.~(\ref{solbend}) approximately by $\int_0^\infty dr GM(r)/r^3$, independent of $b$; with a simple integration by parts using $dM(r)/dr = 4\pi\rho(r)r^2$, where 
$\rho(r)$ is the mass density, gives
\beq
  \langle \theta\rangle \sim -\frac{\pi^2\langle r_0\rangle G}{2\gamma v^2}\int_0^\infty \rho(r) dr.
  \label{estimate}
\eeq
The density in the Sun falls very approximately as $\rho(r) = \rho_c(1-r/R^*)$ where  $\rho_c$ is the central density, and 
$R^*\sim 0.3R_\odot $.  From the solar model \cite{Bahcall},  $\int dr \rho(r) \simeq 3.6 M_\odot/R_\odot^2$, so that 
\beq
    \langle \theta\rangle \sim  - \left\{\frac{}{}\frac{3\pi}{16}\frac{\langle r_0\rangle}{R_\odot}\frac{R^*}{R_\odot}\frac{\rho_c}{\bar\rho} \right\}\frac{GM_\odot}{\gamma v^2 R_\odot} \simeq -\frac{2.0}{\gamma v^2}\frac{GM_\odot}{ R_\odot},
  \label{estimate}
\eeq
where $\bar \rho$ is the average solar mass density.
This estimate is valid to leading order in $b$; the 20\% difference from the numerical result (\ref{numerical}) arises from negative corrections of relative order $-2(b/R^*)^2\ln (R^*/b)$. 

   A similar calculation can be carried out for neutrinos emitted from a neutron star or supernova.   The  characteristic helicity rotation is $\sim GM/\gamma R$, which 
for 10 MeV scale neutrinos is negligible compared with the magnetic rotation produced even by a neutrino magnetic moment of order that 
estimated in the standard model \cite{numag}.

\begin{figure}[t]
\includegraphics*[width=9.0cm]{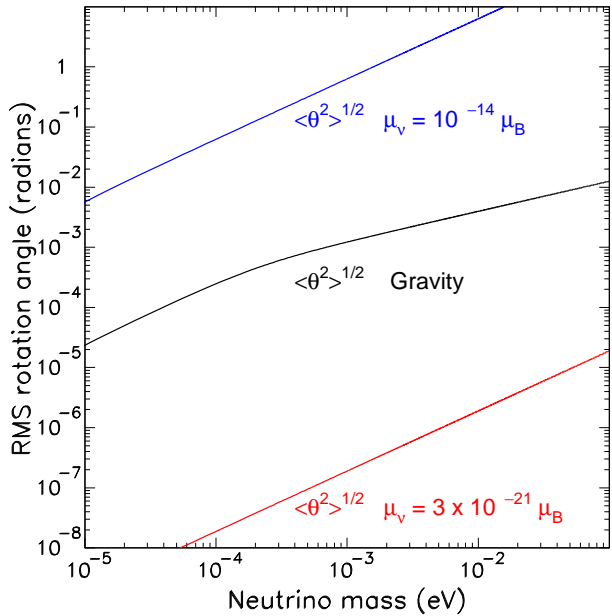}
\caption{Comparison of the root mean square bending angle $\sqrt{\langle \theta^2\rangle}$ of the spin of a primordial neutrino with respect to its momentum from gravitational vs. magnetic effects, as a function of the neutrino mass.  All curves are calculated for the neutrino momentum equal to the temperature. 
The middle curve shows the results of Eq.~(\ref{spinbend10}) for the gravitational bending, for both a Dirac and a Majorana neutrino.  The upper and lower curves are the bending expected from the interaction of a Dirac neutrino magnetic moment, $\mu_\nu$, with a characteristic galactic magnetic field, $\sim 10 \mu$G, for the standard model estimate \cite{fujikawa} of  $\mu_\nu$ (lower curve) \blueflag{with $m_\nu = 10^{-2}$ eV}, and for a magnetic moment $10^{-14} \mu_B$, three orders of magnitude below that which would explain the XENON1T low energy electron events \cite{xenon1t} (upper curve).} 
\label{grav-mag}
\end{figure}

\section{Implications
\label{conclusion}}

   Gravitational perturbations act equally on Dirac and Majorana neutrinos.    As   
relic left-handed Dirac neutrinos are flipped to right-handed, an equal number of right-handed antineutrinos are flipped to left-handed, and since
particles and antiparticles are distinguishable, one could in principle see the depletion experimentally.   On the other hand, if neutrinos are Majorana, the reduction in left-handed neutrinos would not be observable, since the produced left-handed antineutrinos could not be distinguished experimentally from left-handed neutrinos.  

    An initially negative helicity relic neutrino after travelling past the gravitational inhomogeneities in the universe, would have a probability now of being measured with positive helicity,
$P_f=\langle \sin^2(\theta/2)\rangle$.   For a presently relativistic neutrino, with mass less than $10^{-4}$ eV, the flipping probability is
 $\sim  6\times 10^{-7}$.    Since the heaviest neutrino has a mass at least 50 meV \cite{masses}, scattering from density fluctuations should lead, as one sees from Fig~\ref{nugrav_f}, to a population of right-handed relic neutrinos and left-handed relic antineutrinos approaching one in $10^5$.  This effect is too small to be seen in planned experiments to detect relic neutrinos \cite{ptolemy,long} via inverse tritium decay reaction \cite{weinberg}, but it is not beyond the range of eventual measurability.

     Earlier \cite{numag}, we estimated that the bending of the spin of a Dirac neutrino with a diagonal magnetic moment $\mu_\nu$, as it travels through a galaxy, is of order
  \beq
    \langle \theta^2\rangle_{g} \simeq \left(\frac{\mu_\nu B_g }{v}\right)^2  \ell_g \Lambda_g,
    \label{galrot}
\eeq
where $B$ is the average galactic magnetic field, $\ell_g$ is a 
mean crossing distance of the galaxy, $\Lambda_g$ is the characteristic coherence length of the field, and $\mu_B$ is the Bohr magneton.   Unlike gravitational spin bending, the spins of Majorana neutrinos would not be affected by magnetic fields since Majorana neutrinos can have only transition magnetic moments, and the interactions with slowly varying astrophysical magnetic fields cannot change the neutrino mass.

    Equations~(\ref{thetamde}) and (\ref{galrot}) indicate that the scale of spin bending of a non-relativistic thermal neutrino of mass $m_\nu = 10^{-2}$eV by density fluctuations is comparable to that produced by a galactic magnetic field $\sim 10 \mu$G, with $\Lambda_g \sim$ 1kpc and $\ell_g \sim$ 16 kpc,  if the neutrino has a magnetic moment $\mu_\nu \sim 5\times10^{-18}$.    As we see in Fig.~\ref{grav-mag}, the scale of gravitational bending of a neutrino spin with respect to its momentum is well above the magnetic bending produced by the standard model estimate of the magnetic moment \cite{marciano,benlee,fujikawa}, $\sim 3\times 10^{-21} m_{-2}\mu_B$, \blueflag{where $m_{-2}$ is the neutrino mass in units of $10^{-2}$ eV}, but well below that produced by a magnetic moment $1.4-2.9 \times10^{-11}\mu_B$ that would explain the excess of low energy electron events in the XENON1T experiment \cite{xenon1t}.  \blueflag{See discussion in Ref. \cite{numag}.}
   
    Quite generally, neutrino helicity modification, although not measurable by current experiment, is a potentially important probe of cosmic gravitational fields, as well as the interiors of compact objects including the sun, neutron stars, and supernovae.

\acknowledgments
   This research was supported in part by NSF Grant PHY18-22502.  We thank Jessie Shelton, Gil Holder, Stu Shapiro, and Michael Turner for helpful discussions.

\begin{appendix}

\section{Bending of momenta and spins in weak gravitational fields}
\label{gr}

  In this Appendix we summarize the derivations of Eqs.~(\ref{mombend}) and (\ref{spinbend}) for the bending of the momentum and spin in a weak gravitational potential, including the expansion of the universe in the metric, Eq.~(\ref{metricphi}).
  
     The equation of motion of a particle with proper velocity $U^\mu\equiv dx^\mu/d\tau$, where $\tau$ is the proper time of the particle,  propagating through a general gravitational field, is given by the geodesic equation,
\beq
  \frac{dU^\mu}{d\tau}  +\Gamma^{\mu}_{\alpha\beta}U^\alpha U^\beta = 0, 
  \label{geodes}
\eeq
where  $\Gamma^{\mu}_{\alpha\beta} = \frac12 g^{\mu\nu}\left(\partial_\beta g_{\nu\alpha}+ \partial_\alpha g_{\nu\beta}
 -\partial_\nu g_{\alpha\beta}\right)$ is the affine connection.    Using the explicit components of the affine connection for the metric (\ref{metricphi}),\footnote{
 The non-vanishing components of the affine connection 
 are  $\Gamma^i_{00} =\Gamma^0_{i0}= \nabla_i\Phi$, 
 $\Gamma^i_{jk} =-\nabla_k\Phi \delta^i_j  -\nabla_j\Phi \delta^i_k +\nabla_i\Phi\delta^jk$, $\Gamma^0_{00} = a^{-1}da/dx^0$,
  $\Gamma^i_{j0} =\Gamma^i_{0j} = \delta^i_j a^{-1}da/dx^0$, and  $\Gamma^0_{ij} = \delta_{ij}(1-4\Phi)a^{-1}da/dx^0 $.
  }
we see that the spatial velocity, $\vec U^i$, obeys
\beq
    \frac{dU^i}{d\tau} = -\nabla_i \Phi\left((U^0)^2 +(\vec U\,)^2\right) + 2U^i (\vec U\cdot\vec\nabla)\Phi  \nonumber\\ -\frac{2}{a}\frac{da}{d\tau}U^iU^0. 
   \label{dUdtau}
\eeq  
For acceleration along $\vec U$, the  second term on the first line changes the $(U^0)^2 +(\vec U\,)^2$ to $(U^0)^2 -(\vec U\,)^2$ which equals 
$1/a^2$ to zeroth order in $\Phi$; thus  $d(a^2U^i)/d\tau = -\nabla^i\Phi$ along $\vec U$.

    The four-momentum $p_\mu = mg_{\mu\nu}U^\nu$ in general obeys 
\beq
    \frac{dp_\mu}{d\tau}&=& m\frac{dg_{\mu\nu}}{d\tau}U^\nu+mg_{\mu\nu}\frac{dU^\nu}{d\tau} 
     \nonumber\\
    &=& \frac{m}2 \left(\partial_\mu g_{\alpha\beta}\right) U^\alpha U^\beta, 
\eeq
where to find the second line we use $dA/d\tau = U^\mu dA/dx^\mu$, for a function $A$, as well as the geodesic equation combined with the definition of the affine connection.
In the weak field metric with expansion (\ref{metricphi}), the spatial momentum $p_i$ thus obeys
\beq
     \frac{dp_i}{d\tau}&=& \frac{m}{2 }\left(\partial_i g_{\alpha\beta}\right) U^\alpha U^\beta = -ma^2\nabla_i\Phi ((U^0)^2 +(U^i)^2). \nonumber\\
\eeq
Since $dt/d\tau = \gamma$ to zeroth order in $\Phi$,  we find, with expansion,
\beq
  \frac{1}{|\vec p\,|}\frac{d\vec p\,}{dt} = -\left(\frac{1}{v} +v\right)\vec\nabla \Phi, 
  \label{accel}
\eeq  
where $\vec v=d\vec x/du$.  Equation~(\ref{mombend}) follows immediately.  Similarly, in the metric (\ref{metricphi}) [by definition, $p_0<0$],
\beq
    \frac{dp_0}{d\tau}&=& \frac{m}2 \left(\partial_0 g_{\alpha\beta}\right) U^\alpha U^\beta  = -\frac{m}{a} \frac{\partial a}{\partial x^0},
\eeq
since $ g_{\alpha\beta} U^\alpha U^\beta = -1$.   Thus $p_0a$ is conserved.  

     We turn now to spin precession.\footnote{The spin motion was earlier analyzed for a general static metric in Ref.~\cite{voronov} in terms of the tetrad formalism, and for a Dirac particle in Ref.~\cite{silenko} using a Foldy-Wouthuysen transformation of the Dirac equation. .} 
    The helicity is defined in terms of the spin, $\vec S$, in the local Lorentz frame at rest with respect to the particle.   In this frame $S^0 \equiv 0$.  
To determine the equation of motion for $\vec S$, we begin with the spin $\tilde S^\mu$ in the local Lorentz frame at rest in the ``lab,"   which 
obeys the normalization condition, ${\tilde S}_\mu{\tilde S}^\mu = \vec S\,^2$,
and relate $\tilde S^\mu$ to the spin in the weak field metric, denoted here by $\Sigma^\mu$.   

 The normalization condition on ${\Sigma}^\mu$ is
\beq
   {\Sigma}_\mu{\Sigma}^\mu &=& -a^2(1+2\Phi) ({\Sigma}^0)^2 +a^2 (1-2\Phi) \vec {\Sigma}^2=\vec S\,^2. \nonumber\\
       \label{smusmu}
 \eeq
 Thus to first order in $\Phi$,
 \beq
    \tilde S^i = a(1-\Phi){\Sigma}^i,\quad  \tilde S^0 =a (1+\Phi) {\Sigma}^0.    
    \label{SSPhi}
 \eeq
 In addition, ${\Sigma}_\mu U^\mu = 0$, to guarantee that the spin in the particle rest frame has no time component.
 
    The particle spin in the weak field metric obeys the geodesic equation 
\beq
  \frac{d{\Sigma}^\mu}{d\tau}  +\Gamma^{\mu}_{\alpha\beta}{\Sigma}^\alpha U^\beta = 0,
  \label{geodespin}
\eeq
and thus  
\beq
  \frac{d\vec{\Sigma}}{d\tau} &=& -2\vec\nabla\Phi (\vec{\Sigma}\cdot \vec U)  + (\vec U\cdot\vec\nabla\Phi) \vec { \Sigma}
   +  (\vec  {\Sigma}\cdot\vec\nabla\Phi) \vec U \nonumber\\
  && -\frac{1}{a}\frac{da}{dx^0}(U^0 \vec {\Sigma} + {\Sigma}^0 \vec U).    \label{calspini}
\eeq
Equation~(\ref{SSPhi}) implies that to order $\Phi$  the component of the equation of motion of $\vec{\tilde S}$ transverse to $\vec U$ obeys
\beq
   \frac{d\vec{\tilde S}}{d\tau}\Big|_\perp =  \frac{d}{d\tau}\left(a(1-\Phi)\vec{\Sigma}\right)\Big|_\perp& =&  -2\vec\nabla_\perp\Phi (\vec{\tilde S}\cdot \vec U).    \nonumber\\
      \label{spini}
\eeq
Equivalently, $d\vec{\tilde S}/dt |_\perp =  -2\vec\nabla_\perp\Phi (\vec{\tilde S}\cdot \vec v)$, which combined with Eq.~(\ref{accel}) shows that for a massless particle, the spin direction in the lab Lorentz frame remains parallel (or anti-parallel) to the momentum.

  At this stage we transform back to the local Lorentz frame at rest with respect to the particle.  Since $S^0\equiv 0$, the spins in the two Lorentz frames are related by,
\beq
    \vec{\tilde S} = \vec S + (\tilde\gamma -1) \hat v (\hat v\cdot\vec S),
 \eeq
where $\tilde\gamma = (1-\tilde v^2)^{-1/2}$, with the velocity difference of the two Lorentz frames given by $\vec{\tilde v} = [(1+\Phi)/(1-\Phi)] \vec v$.  In components parallel and perpendicular 
to $\vec v$,  $\tilde S_\perp = S_\perp$, and $\tilde S_\parallel = \gamma S_\parallel$.  Thus
\beq
       \frac{d\vec{S}}{d\tau}\Big|_\perp -  \frac{d\vec {\tilde S}}{d\tau}\Big|_\perp   &=& 
      - (\tilde\gamma -1)  (\hat v\cdot\vec S)\frac{d\hat v}{d\tau}\Big|_\perp.
\eeq 
Since $d\hat v/d\tau$ is first order in $\Phi$, we can neglect the distinction between $\vec{\tilde v}$ and $\vec v$, and find
\beq
       \frac{d\vec{S}}{d\tau}\Big|_\perp -  \frac{d\vec {\tilde S}}{d\tau}\Big|_\perp   &=& 
       -\frac{\vec S\cdot \vec U}{(\gamma+1)} \frac{d\vec  U}{d\tau}\Big|_\perp \nonumber\\
       &=& \frac{1}{\gamma+1} \left(\vec S\times \left(\vec U \times \frac{d\vec U}{d\tau}\right)\right)_\perp.  \nonumber\\
\eeq 
The latter term is simply the Thomas precession, at lab frequency $\omega_{\rm Th} = (\gamma^2/(\gamma +1)) \vec v \times \dot {\vec v}$, of an accelerated particle.   With Eqs.~(\ref{spini}) and (\ref{dUdtau}) we then find
\beq
    \frac{d\vec S}{d\tau}\Big|_\perp = -\frac{2\gamma+1}{\gamma+1} (\vec S\cdot \vec U) \vec\nabla_\perp \Phi,
\eeq
from which Eq.~(\ref{spinbend}) follows.

  Equivalently,
\beq
    \frac{d\vec S}{dt}\Big|_\perp = \frac{2\gamma+1}{\gamma+1}\left(\vec S\times(\vec v \times \vec\nabla \Phi)\right)\Big|_\perp,
    \label{spinrot}
\eeq
indicating that the spin feels an effective velocity-dependent torque  $(\mu \vec B)_{\rm eff} = [(2\gamma+1)/2(\gamma+1)]( \vec v \times \vec\nabla \Phi$).   The non-relativistic limit of this equation gives Schiff's result for precession of a spin in the Gravity Probe B experiment  \cite{schiff} (see also Ref.~\cite{weinbergGR}), while in the fully relativistic limit, $\gamma\to\infty$, the spin remains at the same angle with respect to the momentum.

\section{Gravitational spin rotation of neutrinos emitted from a spherical body}

  We detail here the calculation of the relative spin rotation of neutrinos emitted from a spherical star, applicable to solar neutrinos as well as neutrinos from supernovae and neutron stars.   We first convert the $z$ integral in Eq.~(\ref{solbend}), with $z_0$ set to 0, to an integral over $r$, so that 
\beq
   \gamma v^2 \theta(b)  &=& -b\int_{b}^\infty dr \, \frac{G M(r)}{r^2\sqrt{r^2-b^2}},   
    \label{solbend11}
\eeq 
Then we average the neutrino emission over the stellar volume with a spherically symmetric normalized spatial emission probability $p_\nu(r_0)d^3r_0$, in terms of cylindrical coordinates ($d^3r_0 = 2\pi b db \, dz$),
\beq
  \langle \theta(b) \rangle &=&\int 2\pi b db \, dz\, \int dr_0 p_\nu(r_0)\delta\left(r_0-\sqrt{b^2+z^2}\right) \theta(b),\nonumber\\
   &=& \int_0^{R_\odot} 4\pi r_0 dr_0 p_\nu(r_0) \int_0^{r_0} \frac{b\,db}{\sqrt{r_0^2-b^2}} \theta(b),
\eeq
where in the first line the ranges of the $b$ and $z$ integrals are constrained by the delta function.
Thus 
 \beq
   \gamma v^2 \langle\theta\rangle  &=& - \int_0^{R_\odot} 4\pi r_0 dr_0 p_\nu(r_0) \int_0^{r_0} \frac{b^2db}{\sqrt{r_0^2-b^2}}\nonumber\\ &&\hspace{24pt}\times\int_{b}^\infty dr \, \frac{G M(r)}{r^2\sqrt{r^2-b^2}}.
 \label{solbend2}
\eeq  
Interchanging the order of the $r$ and $b$ integrals, we see that their product is equivalent to
\beq
   && \int_0^\infty dr\frac{GM(r)}{r^2} f(r,r_0), 
\eeq
where
\beq
  f(r,r_0) = \Theta(r_0-r) r W(r_0/r)  + \Theta(r-r_0) r_0 W(r/r_0)  \nonumber
\eeq 
with
\beq
   W(\xi) &= &\int_0^1 \frac{x^2\,dx}{\sqrt{1-x^2}\sqrt{\xi^2-x^2}} \nonumber\\
&=&
 \int_0^1 dx\frac{\sqrt{1-x^2}}{\sqrt{\xi^2-1+x^2}}, \quad \xi>1. 
\eeq
Equation~(\ref{solbend3}) follows directly.


\end{appendix}


\begin{thebibliography}{99}
\bibitem{masses} The best-value fits to the mass squared differences, $\Delta m^2_{ij} \equiv m^2_i - m^2_j$, are \cite{esteban}
$\Delta m^2_{21}  =  (7.50 + 0.19 - 0.17) \times 10^{-5} {\rm eV}^2, 
\Delta m^2_{31,N}  =  (2.524 + 0.039 - 0.040)\times 10^{-3} {\rm eV}^2, 
\Delta m^2_{31,I}  =  (-2.514 + 0.038 - 0.041)\times 10^{-3} {\rm eV}^2$.
\end{thebibliography}
\end{document}